\documentclass[usenatbib]{mn2e}
\usepackage{times}
\usepackage{epsfig}
\usepackage{url}

\newcommand*{\hnull}{\ensuremath{H_{0}}}
\newcommand*{\lnull}{\ensuremath{\lambda_{0}}}
\newcommand*{\onull}{\ensuremath{\Omega_{0}}}
\newcommand*{\etc}{etc.} 
\newcommand*{\eg}{e.g.}
\newcommand*{\ie}{i.e.}
\newcommand*{\percent}{per cent} 
\newcommand*{\perse}{per se}
\newcommand*{\via}{via}

\newcommand*{\ack}{Acknowledgements} 
\newcommand*{\pmzd}{parametrized} 
\newcommand*{\pmzn}{parametrization}
\newcommand*{\bestfit}{best-fitting}
\newcommand*{\Fig}{Fig.}
\newcommand*{\Figs}{Figs.}
\newcommand*{\Sect}{Section}
\newcommand*{\eqn}{equation}
\newcommand*{\fig}{fig.}
\newcommand*{\figs}{figs.}
\newcommand*{\sect}{section}
\newcommand*{\ndash}{--} 
\newcommand*{\pdash}{~--~} 
\newcommand*{\kdash}{ -- } 
\newcommand*{\software}[1]{{\sc #1}}

\title[Supernovae, inhomogeneity and dark matter]%
{The $m$--$z$ relation for Type~Ia supernovae, locally inhomogeneous 
cosmological models, and the nature of dark matter}

\author[P.~Helbig]%
{Phillip Helbig\thanks{E-mail: helbig@astro.multivax.de}\\
Thomas-Mann-Str.~9, D-63477 Maintal, Germany}

\begin{document}

\date{Accepted 2015 May 11.  Received 2015 May 9; in original form 2015 
February 15}

\pagerange{\pageref{firstpage}--\pageref{lastpage}} \pubyear{2015}

\maketitle
\label{firstpage}

\begin{abstract}
The $m$--$z$ relation for Type~Ia supernovae is one of the key pieces of
evidence supporting the cosmological `concordance model' with $\lnull
\approx 0.7$ and $\onull \approx 0.3$.  However, it is well known that
the $m$--$z$ relation depends not only on \lnull\ and \onull\ (with
\hnull\ as a scale factor) but also on the density of matter along the
line of sight, which is not necessarily the same as the large-scale
density.  I investigate to what extent the measurement of \lnull\ and
\onull\ depends on this density when it is characterized by the
parameter $\eta$ ($0 \le \eta \le 1$), which describes the ratio of
density along the line of sight to the overall density.  I also discuss
what constraints can be placed on $\eta$, both with and without
constraints on \lnull\ and \onull\ in addition to those from the $m$--$z$
relation for Type~Ia supernovae. 
\end{abstract} 

\begin{keywords}
supernovae: general\kdash cosmological parameters\kdash 
cosmology: theory\kdash dark energy \kdash dark matter.
\end{keywords}

\section{Introduction}
\label{intro}

In the last 15 years or so, cosmological observations have improved
greatly and it also appears that the values are converging on their true
values.\footnote{The case for the `concordance model' with $\lnull
\approx 0.7$ and $\onull \approx 0.3$ was already made by 
\citet{JOstrikerPSteinhardt95a}; 
the values of the concordance model thus do not need the supernova data,
though of course adding more data improves the constraints.  While the
corresponding uncertainties have dramatically decreased 
\citep[\eg][]{EKomatsuetal2011a,PLANCKXVI2014a}, 
the values themselves have remained constant over the last twenty
years.} This allows us to answer such questions (provided, of course,
that `standard assumptions' hold) as whether the Universe will expand
for ever (yes), how old it is, whether it is accelerating now (yes),
when it started accelerating, \etc\  (With the assumption of a simple
topology, the Universe is finite if $\lnull + \onull  > 1$, but since
observations indicate that this value is very close to 1, we cannot yet
answer this question.)  Among the most important of these observations
are those by the Supernova Cosmology Project 
\citep[\eg][]{AGoobarSPerlmutter95a,SPerlmutteretal95a,
SPerlmutteretal98a,SPerlmutteretal99a,RAmanullahetal2010a,NSuzukietal2012a}
and the High-$z$ Supernova Search team 
\citep[\eg][]{PGarnavichetal1998a,ARiessetal98a,ARiessetal2000a}
(see also the reviews by 
\citealt{ARiess2000a}, 
\citealt{BLeibundgut2001a,BLeibundgut2008a},
and 
\citealt{AGoobarBLeibundgut2011a})
which provide joint constraints on \lnull\ and \onull.  Combined with
other observations 
\citep[\eg][]{EKomatsuetal2011a,PLANCKXVI2014a}, 
these lead to quite well constrained values for the cosmological
parameters 
\citep[\eg~\fig~5 in][]{NSuzukietal2012a}.  
Although the supernova data alone allow a relatively wide range of
significantly different other models, it is interesting that the
\bestfit\ values obtained from these observations using the current data
are quite close to the much better constrained values using combinations
of several observations without the supernova data, at least under the
assumptions with which the former were calculated.  However, since the
$m$--$z$ relation depends not only on \lnull\ and \onull\ (with \hnull\
as a scale factor) but also on the distribution of matter along and near
the line of sight, the dependence of conclusions drawn from the $m$--$z$
relation for Type~Ia supernovae on this matter distribution should be
investigated.  Alternatively, these observations can perhaps tell us
something about this distribution. 

The plan of this paper is as follows.  \Sect~\ref{theory} sketches the
basic theory used in this paper.  In \sect~\ref{history} I briefly
review previous investigations of the influence of a locally
inhomogeneous universe on the $m$--$z$ relation.  \Sect~\ref{main}
describes the calculations done and discusses the results.  Summary,
conclusions and outlook are presented in \sect~\ref{final}.

\section{Basic theory}
\label{theory}

\citet*[hereafter KHS]{RKayserHS97Ra} 
developed a general and practical method for calculating cosmological
distances in the case of a locally inhomogeneous universe.  See KHS for
details (and for a description of the notation, which is followed here);
here I repeat only the most important points for the purpose of this
paper. 

If the Universe is homogeneous, then the fact that light propagates
along null geodesics provides sufficient information to calculate
distances from redshift.  If the Universe is locally inhomogeneous, then
distances which depend on angular observables related to the propagation
of radiation will differ from the homogeneous case because more or less
convergence will change the angle involved (the angle at the observer in
the case of the angular-size distance, that at the source in the case of
the luminosity distance).  The basic idea is that one considers a
Universe which is homogeneous and isotropic on large scales, this
determining the global dynamics via the Friedmann\ndash Lema\^{\i}tre
equation.  Local inhomogeneities are modelled as clumps, where the extra
matter in the clumps is taken from the surrounding matter.  Thus, a beam
which propagates between clumps will have only this thinned-out matter
inside the beam, while outside the beam the average density (taking both
the thinned-out background matter and the clumps into account) is
approximately equal to the global density (precisely so in the limit of
an infinitesimal beam). 
\citet{YZeldovich64a},
\citet{VDashevskiiYZeldovich65a} 
and 
\citet{VDashevskiiVSlysh66a}
developed a general differential equation for the distance $l$ between
two light rays on the boundary of a small light cone (the beam)
propagating far away from all clumps of matter in a locally
inhomogeneous universe: 
\begin{equation}\label{zds-g}
\ddot{l} = -4\upi G\eta\rho \,l + {\dot{R}\over R}\,\dot{l}\quad ,
\end{equation}
where $G$ is the gravitational constant, $R$ the scale factor, and
$\eta$ (defined below) and the density $\rho$ are functions of time (a
dot indicates differentiation with respect to time).  The first term can
be interpreted as Ricci focusing due to the matter inside the beam, and
the second term is due to the expansion of space during the light
propagation.  The key assumption here is that while the the density
$\rho$ within the beam can differ from the overall density, the overall
dynamics of the universe is still described by the Friedmann\ndash
Lema\^{\i}tre equation.  The assumption that the light propagates far
from all clumps means that Weyl focusing (shear) can be neglected.  In
the case that the densities inside the beam and outside the beam are the
same, one of course recovers the homogeneous case. 

Since the angular-size distance is defined as $D = l/\theta$, where
$\theta$ is the angle at the apex of the beam (at the observer, not at
the observed object), $D$ follows the same differential equation as $l$.
 Making use of this, one can derive a general equation for the
angular-size distance, valid for all (perturbed, in the sense described
above) Friedmann\ndash Lema\^{\i}tre cosmological models and all
reasonable (see below) values of $\eta$.  KHS described the
inhomogeneity \via\ the parameter $0 \le \eta \le 1$, where $\eta$ is
ratio of the density inside the beam to the global density or,
alternatively, the fraction of matter which is homogeneously
distributed, as opposed to being clumped.\footnote{This is sometimes
denoted by $\alpha$.  I, and some others, use $\eta$ because locally
inhomogeneous cosmological models are often used in gravitational
lensing (which \perse\ implies local inhomogeneities) where $\alpha$ is
almost always used to denote the deflection angle.} This  leads to a
second-order differential equation for the angular-size distance
(\eqn~(33) in KHS) which can be efficiently integrated numerically: 
\begin{equation}\label{RKeqG}
Q\, D'' + \left({2Q\over 1+z} + {1\over 2}\, Q'\right)\,D'
      + {3\over 2}\,\eta\,\Omega_0 (1+z)\, D = 0 \quad ,
\end{equation}
where
\begin{equation}\label{q-g}
Q(z) = \Omega_{0}(1 + z)^{3} -
       (\Omega_{0} + \lambda_{0} -1)(1 + z)^{2} +
       \lambda_{0} \quad.
\end{equation}
In the locally inhomogeneous case as well the luminosity distance, which
is needed in this paper, is larger than the angular-size distance by a
factor of $(1+z)^{2}$. 

This change, compared to the perfectly homogeneous case, is essentially
a negative gravitational-lensing effect.  In a conventional
gravitational-lensing scenario, if the density at a given redshift
between two light rays is higher than the overall density (the
corresponding overdensity being `the lens'), then there will be more
convergence than in the case where the two densities are the same.  In
the case of light propagating between clumps, as described above, the
situation is reversed, and the density between the light rays defining
the distance-related angle is less than the overall density.  This means
that there is (negative) Ricci focusing (and no Weyl focusing), making
objects appear fainter than they would be in the completely homogeneous
case.  Of course, this is only a rough model, but can be expected to be
more realistic than the completely homogeneous case and to determine not
just the sign of the difference but also give at least an estimate of
its strength. 

Obviously, one cannot have $\eta < 0$.  However, it does not make sense
to have $\eta > 1$ either.  While it is certainly possible that the
average density inside the beam could be greater than the global
density, such cases are either unrealistic or not useful. The limiting
case where the density in the beam is greater than the global density by
a constant factor at every redshift is unrealistic because this would
imply the existence of overdense regions with an extreme length-to-width
ratio which are aligned between us and the source, which is incompatible
with homogeneity and isotropy on large scales and would also put us in a
special position.  The other limiting case where a single compact object
increases the density in the beam to above the global density is
certainly possible, but observationally would show up as a
gravitational-lens effect and should be analysed as such (perhaps by
adopting $\eta \approx 0$ for the distance calculation and explicitly
calculating the amplification).  Of course, cases between these two
extremes are possible, but it is clear that $\eta$ must be between 0 and
1 if it is used as an additional parameter in the manner described by
KHS; lines of sight which, due to fluctuations, are slightly denser than
the overall density are certainly possible, but are not usefully \pmzd\
by $\eta$ in the style of KHS.  (But see 
\citealt*{JLimaVS14a} 
for a toy model with an interesting extension of the $\eta$ concept.) 
Note that $\eta$ does not have to be constant as a function of redshift,
and the code described in KHS supports an arbitrary dependence of $\eta$
on $z$.  Also, it could be different for different lines of sight.  It
was pointed out by 
\citet{SWeinberg76a} 
that $\eta$ must be 1 when averaged over all lines of sight (allowing
for the moment higher-than-global densities to be \pmzd\ by $\eta > 1$),
which follows from flux conservation.  However, in practice lines of
sight will probably avoid concentrations of matter, due to selection
effects or design: distant objects will be more difficult to observed if
there is luminous matter along the line of sight or if there is
absorbing matter along the line of sight.\footnote{Matter along the line
of sight can increase the apparent brightness and thus make objects
visible which otherwise would not be.  This phenomenon, known as
`amplification bias' in gravitational lensing, is relevant only if the
luminosity function is steep enough (since otherwise the magnification
of the area of sky observed, which reduces the number of objects per
observed area, will dominate, resulting in fewer objects in a
flux-limited sample).  However, the whole point of the $m$--$z$ relation
for Type~Ia supernovae is that they are standard candles, or can be
adjusted to behave as standard candles with the help of other
observations, which means that the differential (integral) luminosity
function is essentially a delta (Heaviside) function, so the
amplification bias plays no role here.  Also, since objects much fainter
than supernovae can be detected in the corresponding observations, no
realistic amplification would make an otherwise undetectable object
visible.}  If these selection effects do not exist, and if the sample is
large enough, then the `Safety in Numbers' effect 
\citep{DHolzELinder05a} 
allows one to effectively assume $\eta = 1$, with inhomogeneity merely
increasing the dispersion, roughly linearly with redshift.  However, 
\citet{CClarksonetal12a} 
point out that most narrow-beam lines of sight are significantly
underdense, even for beams much thicker than those considered in this
paper.  (On the other hand, they also point out that this does not
necessarily lead to a reduction in brightness if one drops the
assumption that inhomogeneities can be modelled as perturbations on a
uniformly expanding background, a point also emphasized by 
\citealt{KBolejkoPFerreira12a}; 
see also
\citealt{SBagheriDSchwarz14a}.)

The situation discussed above corresponds to the situation where the
beam contains a density $\eta$ times the global density at a given
redshift and outside the beam the density is equal to the global
density.  In practice, this means that a fraction $\eta$ of the mass in
the Universe is smoothly distributed and a fraction $1 - \eta$ is
contained in clumps outside the beam.  Of course, `smoothly' depends on
the size of the beam; for example, small objects are part of the
`smooth' component, not only in the limiting case where the smooth
component consists of free elementary particles.  The important point is
that their angular size is small compared to that of the beam.  $\eta$
is thus also a function of angle: the larger the angle, the more
representative is the matter within the beam, so that $\eta$ approaches
1 for large enough angles.  Since the beams of supernovae at
cosmological distances are extremely thin objects (the thinnest objects
ever studied by science), evidence for $\eta < 1$ should be most obvious
in the $m$--$z$ relation for Type~Ia supernovae. 

A given value of $\eta$ along a given line of sight does not imply that
this value does not change along the line of sight, although that is of
course a possibility, but rather that the influence on angle-dependent
distances can be described by an effective value of $\eta$ which is some
appropriate average of a value which varies along the line of sight.
This means that it is possible for the density along the line of sight
to be larger than the global density at some points, but this is not in
contrast with the claim above that $\eta > 1$ is not useful as long as
the \emph{effective} value $\eta_{\mathrm{eff}} \leq 1$.  Another
complication is that essentially all lines of sight to supernovae will
have a density higher than the globally average cosmological density due
to the overdensities associated with the Milky Way and with the
supernova host galaxy (and corresponding clusters).\footnote{I thank
Philip Bull for first pointing this out to me.} However, since the
absolute magnitudes of supernovae are not known from first principles,
but rather calibrated from observations, this effect is, to a first
approximation, unobservable, since it is essentially a renormalization
of the absolute magnitude.  Even if this extra matter associated with
the galaxies at the ends of the beam would increase the density inside
the beam to larger than the global density, it is not useful to think of
this as $\eta > 1$, since I want to compare the standard assumption
(completely homogeneous Universe, at least as far as light propagation
is concerned) with that of a more realistic distribution.  The point of
comparison, the $m$--$z$ relation for a homogeneous Universe, also
contains extra matter at each end of the beam, and hence extra
convergence.  As far as I know, no-one has ever taken this into account
and it is not necessary if one is interested only in the differences. 
(This would have to be taken into account, though, if the absolute
magnitude of objects at cosmological distances were known independently
of observation.) 

Although the term `dark matter' suggests something opaque, the defining
characteristic is lack of interaction with electromagnetic radiation.
Thus, not only does dark matter not glow, it is also transparent.  It is
thus irrelevant whether dark-matter objects within the beam
significantly cover a source as seen by an observer.  (Of course, when
comparing observed to calculated brightness, one must correct for
extinction due to `conventional' matter\pdash it can also be dark in the
sense that it does not radiate, but it is not transparent.)  Here, I am
using the term `dark matter' to refer to the `missing matter', \ie~that
responsible for the difference between the density due to baryonic
matter (other non-baryonic but known particles (neutrinos) do not
increase this significantly) and the global density of the Universe, as
measured on large scales.  Of course, non-radiating baryonic matter does
exist, but we know from constraints from big-bang nucleosynthesis that
this cannot be a significant fraction of the missing matter.  This
reflects current usage, \eg~the `DM' in `$\Lambda$CDM', and is more
convenient than `not yet identified non-baryonic matter'. 

Since we know that the Universe is not exactly homogeneous and
isotropic, $\eta \neq 1$ is the most obvious departure from the simplest
cosmological model (the Einstein\ndash de~Sitter model with $\lnull =
0$, $\onull = 1$, and $\eta = 1$, although the last item is often not
stated explicitly), but there is not much literature on this topic.
(There are, though, several recent papers investigating whether `dark
energy' could be something other than the traditional cosmological
constant, \eg~whether the equation of state $w$ differs from $-1$,
whether it changes with time etc, even though there are no observations
which indicate this.  Of course, that does not mean that one should not
look.) 

If $\eta$ is allowed to vary from one line of sight to another, one
could regard this as an additional contribution to the uncertainty in
the distance modulus, much the same as the uncertainty in the absolute
magnitude.  Theoretically, fitting the observations for a constant value
of $\eta$ would result in a worse fit for such cases if this additional
uncertainty is ignored or in a larger allowed region of parameter space
if it is included in the error budget.  With some assumptions, one could
try to take this additional dispersion into account and/or correct for
it; see, \eg, 
\citet*{RAmanullaEMoertsellAGoobar2003a}, 
\citet{CGunnarssonDGJM2006a},
\citet{JJoenssonDGGML2006a},
\citet*{JJoenssonEMoertsellJSollerman2009a}.  
In practice, with a large number of objects and only a few variables,
the difference in goodness of fit is well within the expected range of
values for the case in which $\eta$ is the same along all lines of
sight.  Alternatively, with current data it is also a relatively small
contribution to the error budget.  Thus, if observations suggest $0 <
\eta < 1$, it would be unclear if this is evidence for the corresponding
value of the global value of $\eta$ or whether this is a compromise
between lines of sight with lower and higher values.  However, if
observations indicate $\eta = 0$ or $\eta = 1$, then this would be
evidence for the corresponding global value, because these are the
extreme values of $\eta$ and cannot result from averaging. 

Of course, more complicated models are possible.  In this paper, I
consider only models in which $\eta$ is a constant function of redshift
and the same along all lines of sight\footnote{See 
\citet{CGunnarssonDGJM2006a} 
for a discussion of a $z$-dependent $\eta$ in the context of the $m$--$z$
relation for Type~Ia supernovae.}; also, in all cases but one, it is
independent of the other cosmological parameters.  The variation between
these models, however, is certainly larger than the realistic range of
the possible influence of $\eta$ on the $m$--$z$ relation for Type~Ia
supernovae.

\section{Brief history}
\label{history}

The effects of a locally inhomogeneous universe on quantities important
for observational cosmology were first investigated in a series of
papers by 
\citet{YZeldovich64a},
\citet{VDashevskiiYZeldovich65a} 
and 
\citet{VDashevskiiVSlysh66a}.  
\citet{CDyerRRoeder72a} 
discussed the special case of $\lnull = 0$ but with \onull\ as a free
parameter for $\eta = 0$ (where there is an analytic solution)  and for
general $\eta$ values 
\citep{CDyerRRoeder73a}.  
As a result, the distance for $\eta = 0$ is sometimes referred to as the
Dyer\ndash Roeder distance.  KHS presented a second-order differential
equation and numerical implementation valid for the general case
($-\infty < \lnull < \infty$, $0 \leq \onull \leq \infty$, $0 \leq \eta
\leq 1$).  Kantowski  and collaborators 
\citep*{RKantowski1969a,RKantowskiVB1995a,RKantowski1998a,
RKantowskiKT2000a,RKantowskiRThomas2001a,RKantowski2003a}
have stressed the importance of $\eta$ for the interpretation of the
$m$--$z$ relation for Type~Ia supernovae and have provided numerical
implementations using elliptic integrals for the special values of
$\eta$ of 0, $\frac{2}{3}$, and 1. 
\citet{SPerlmutteretal99a}
considered the effect of $\eta \neq 1$ on their results (see their
\fig~8) and concluded that, at least in the `interesting' region of the
\lnull--\onull\ parameter space, it had a negligible effect 
\citep[see also][]{JJoenssonDGGML2006a}.  
The reason for the current paper is that, with the larger number
of supernovae now available, this is no longer the case.  Further
investigation has often been motivated by the $m$--$z$ relation for
Type~Ia supernovae 
\citep*[\eg][]{MGoliathEMoertselll2000a,EMoertsellAGoobarLBergstroem2001a}.  
It has also been investigated, \via\ comparison with explicit
ray-tracing through mass distributions derived from simulations or
observations, whether $\eta$ is a useful \pmzn\ for local inhomogeneity 
\citep[\eg][]{LBergstroemGGM2000a,EMoertsell2002a}
(and the conclusion is that it is a useful approximation, at least for
cosmological models which are otherwise realistic). 

There seem to be three schools with respect to the attitude taken to the
possible influence of inhomogeneities on cosmological parameters derived
from the $m$--$z$ relation for Type~Ia supernovae.  One school ignores it
completely, assuming a completely homogeneous Universe as far as the
calculation of the luminosity distance is concerned 
\citep[\eg][]{ARiessetal98a}, 
or provides some limited justification for not considering it further 
\citep[\eg][]{MBetouleetal14a}.  
Another school emphasizes that the problem is not completely understood,
the amount of uncertainty is unknown, and even the sign of some effects
is unclear 
\citep[\eg][]{CClarksonetal12a}.  
A third school uses some approximation to at least get an idea of the
size of possible effects 
\citep[\eg][]{EMoertsellAGoobarLBergstroem2001a}.  
(While 
\citealt{SPerlmutteretal99a} 
did consider the possible influence of $\eta$, hence belonging to the
third school, at least at that time, with their data then it was not a
significant source of uncertainty in their main result.  One purpose of
this paper is to show that this is no longer the case.)

\section{Calculations, results and discussion}
\label{main}

I have used the publicly available `Union2.1' sample of supernova data 
\citep{NSuzukietal2012a} 
and calculated $\chi^{2}$ and the associated probability following 
\citet{RAmanullahetal2010a} 
on regularly-spaced grids of various extents and resolutions in the
\lnull--\onull--$\eta$ parameter space.  This \emph{assumes}, of course,
that $\eta$ is a free parameter on the same footing as \lnull\ and
\onull.  My goal is not to obtain the `best' cosmological parameters,
not even the `best' ones from the supernova data alone.  Rather, it is
to investigate the influence of $\eta \neq 1$ on the interpretation of
the $m$--$z$ relation for Type~Ia supernovae. I have thus intentionally
made the supernova data as precise as possible, by using only the
statistical uncertainties (\ie~column 4 in the publicly available data
file) and fixing \hnull\ at 70~kms$^{-1}$Mpc$^{-1}$, which implies $M =
-19.3182761161$.  Thus, all increase in the allowed region of parameter
space (at a given confidence level) is due only to the influence of
$\eta$.\footnote{Since the goal is not to obtain the best constraints on
\lnull\ and \onull, but rather to investigate the influence of $\eta$ on
the constraints, I have retained the Union 2.1 sample with which I began
this investigation, rather than updating it to use, \eg, that used by 
\citet{MBetouleetal14a}.  
Those with better access to such data will always have a better sample
than that which is publicly available. Since even 
\citet{MBetouleetal14a} 
do not consider $\eta$ at all, it is perhaps important at the moment for
a theorist to take a step back for a more general view in order to
contrast with continual updates using somewhat better samples.  It is
important, though, that the Union 2.1 sample is significantly larger
than those used in the early works discussed in \sect~\ref{intro}.} 

I have calculated $\chi^{2}$ and the corresponding probability on two
three-dimensional grids: a larger, lower-resolution grid 
\begin{displaymath}
\begin{array}{@{}r@{\,}c@{\,}c@{\,}c@{\,}rl@{\,}c@{\,}rr}
  -5 & < & \lnull & <  &  5 & \Delta\lnull & = & 0.02 & 
    \mathrm{\hspace{2.3em}(500\ values)} \\
   0 & < & \onull & <  & 10 & \Delta\onull & = & 0.02 & 
    \mathrm{(500\ values)} \\
   0 & < & \eta   & <  &  1 & \Delta\eta   & = & 0.01 & 
    \mathrm{(100\ values)}
\end{array}
\end{displaymath}
and a smaller, higher-resolution grid
\newlength{\bla}
\setlength{\bla}{0.8em}
\begin{displaymath}
\begin{array}{@{\hspace*{\bla}}r@{\,}c@{\,}c@{\,}c@{\,}rl@{\,}c@{\,}rr}
   0 & < & \lnull & < & 1.5 & \Delta\lnull & = & 0.003125 & 
    \mathrm{(480\ values)}\\
   0 & < & \onull & < & 1.0 & \Delta\onull & = & 0.003125 & 
    \mathrm{(320\ values)}\\
   0 & < & \eta   & < & 1.0 & \Delta\eta   & = & 0.01     & 
    \mathrm{(100\ values)}
\end{array}
\end{displaymath}
(This paper contains no plots based on the larger, lower-resolution
grid; the corresponding calculations were done to make sure that there
is no appreciable probability outside of the range of the smaller,
higher-resolution grid.)  Since three-dimensional contours cannot be
fully represented in two dimensions, I present various two- and
one-dimensional visualizations in order to illustrate the influence of
$\eta$. 

All contours in two (three) dimensions have been calculated as the
smallest-area closed curve (smallest-volume closed surface) which
encloses the corresponding fraction of the probability.  I have used the
standard values $0.683$, $0.954$ and $0.997$; these correspond to
1--$\sigma$, 2--$\sigma$ and 3--$\sigma$ in the Gaussian case.  However,
I have made no assumption about Gaussianity, since I have calculated the
contours explicitly, rather than plotting them at the corresponding
fraction of the peak likelihood under the Gaussian assumption.  For all
plots, the area outside of the plot has been assigned a probability of
zero.  Otherwise, no priors other than those explicitly stated have been
used.  In particular, no prior information on the values of the
cosmological parameters from other tests have been used; what I show
depends on the supernova data only. 

\begin{figure}
\epsfig{file=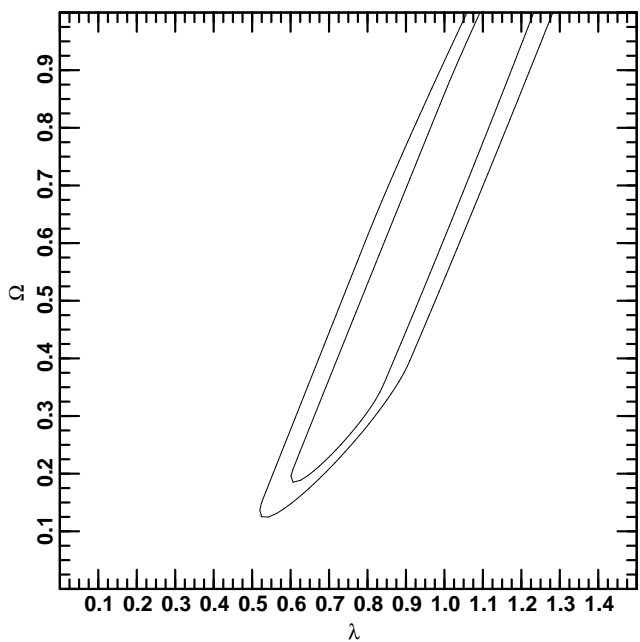,width=0.8\columnwidth}
\caption{Projection of three-dimensional probability distribution
along the $\eta$-axis.} 
\label{closeup-project-lo}
\end{figure}
\Figs~\ref{closeup-project-lo}, 
\begin{figure}
\epsfig{file=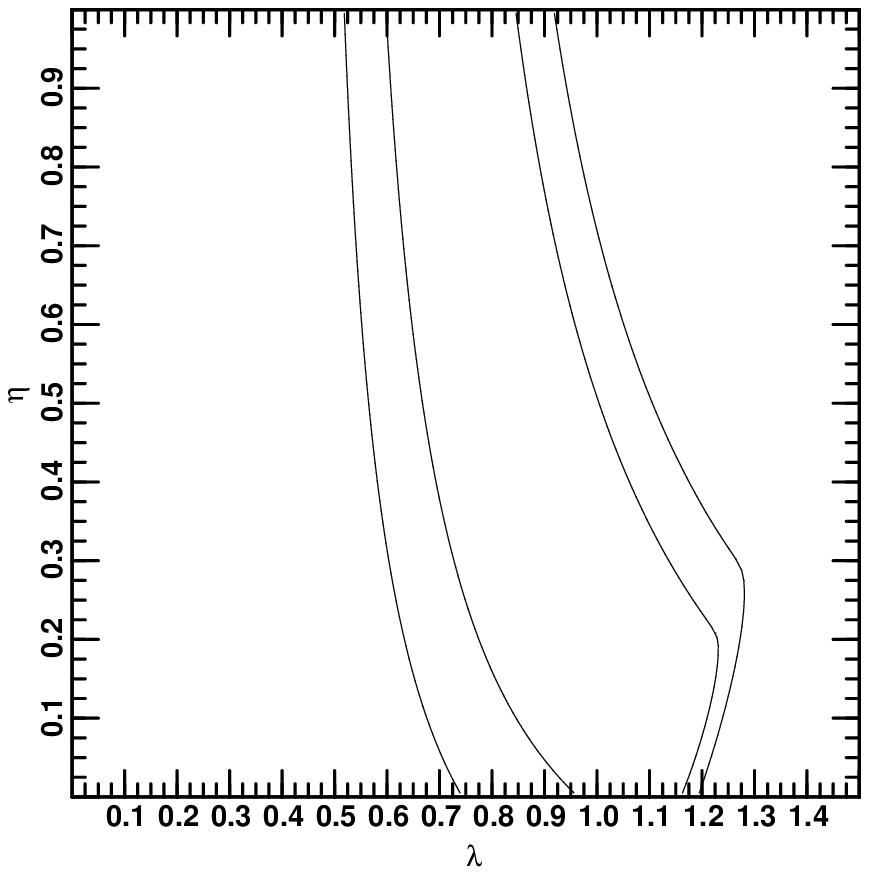,width=0.8\columnwidth}
\caption{Projection of three-dimensional probability distribution
along the \onull-axis.} 
\label{closeup-project-le}
\end{figure}
\ref{closeup-project-le}, and
\begin{figure}
\epsfig{file=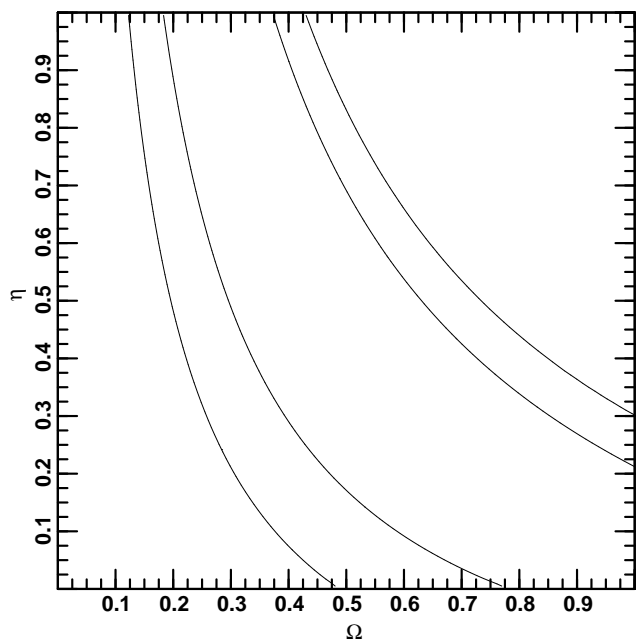,width=0.8\columnwidth}
\caption{Projection of three-dimensional probability distribution
along the \lnull-axis.} 
\label{closeup-project-oe}
\end{figure}
\ref{closeup-project-oe} 
show \emph{projections} of the three-dimensional contours along one axis
on to the plane spanned by the other two axes for the smaller,
higher-resolution grid.  It can be seen that the combination of \lnull\
and \onull\ is well constrained, as are both individually, while $\eta$
is hardly constrained at all.  Note also that \lnull\ and \onull\ are
less constrained for lower values of $\eta$.  (The contours at $0.954$
and $0.997$ cannot be distinguished in these plots.)  The relatively
sharp bend in the lower right contours in \Figs~\ref{closeup-project-lo}
and \ref{closeup-project-le} is due to the fact that I have assigned a
probability of $0$ to models which have no big bang (see the discussion
of \figs~1 and 2 in 
\citet{PHelbig12Ra} 
and references therein for an explanation). 

Another way of visualizing these three-dimensional contours is to make
\emph{cuts} through them for a fixed value of one of the parameters. 
\begin{figure}
\epsfig{file=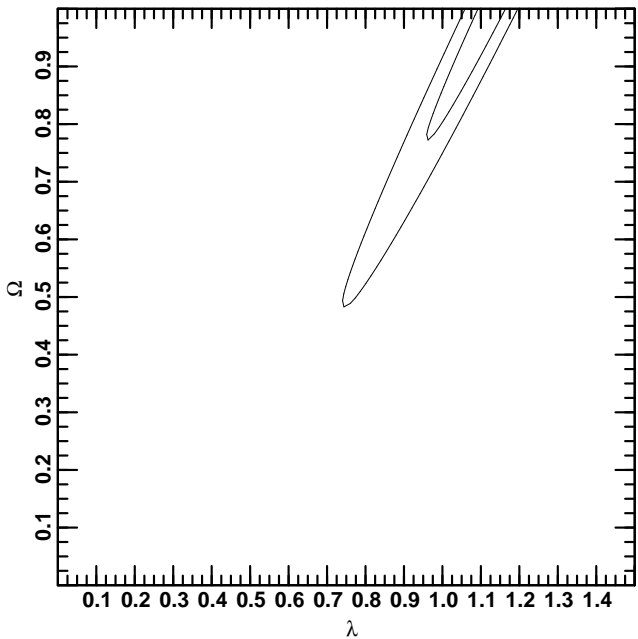,width=0.8\columnwidth}
\caption{Cut through the three-dimensional probability distribution
perpendicular to the $\eta$-axis for $\eta=0.005$.} 
\label{closeup-cut-1-lo}
\end{figure}
\Figs~\ref{closeup-cut-1-lo}, 
\begin{figure}
\epsfig{file=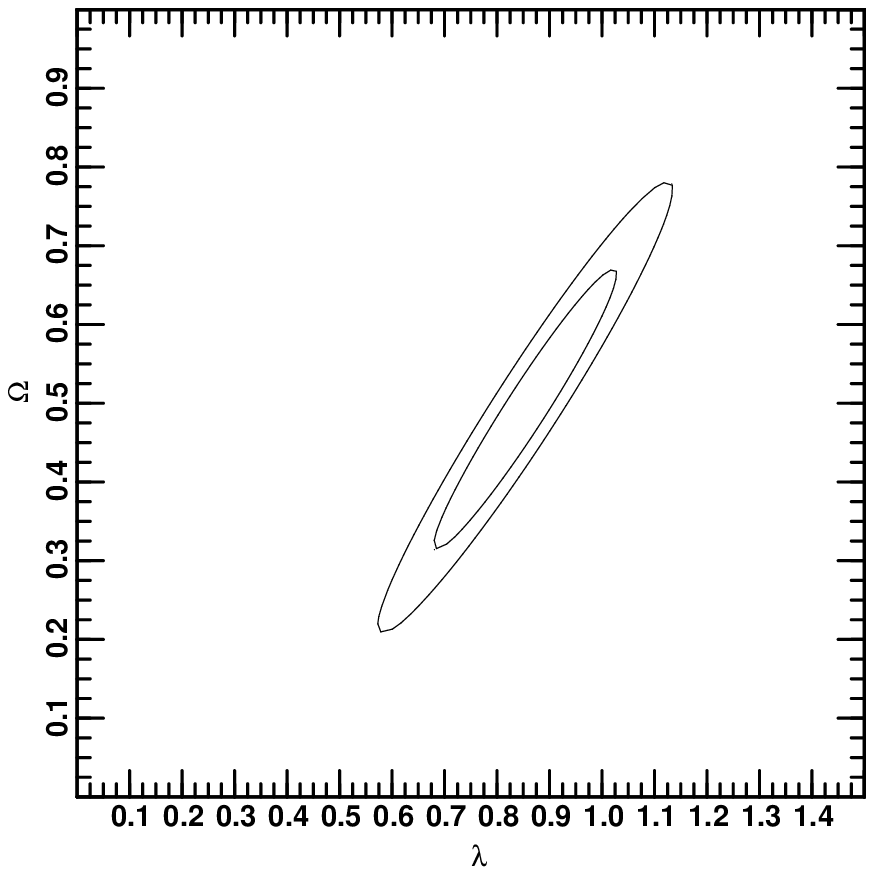,width=0.8\columnwidth}
\caption{Cut through the three-dimensional probability distribution
perpendicular to the $\eta$-axis for $\eta=0.455$.} 
\label{closeup-cut-2-lo}
\end{figure}
\ref{closeup-cut-2-lo}, and
\begin{figure}
\epsfig{file=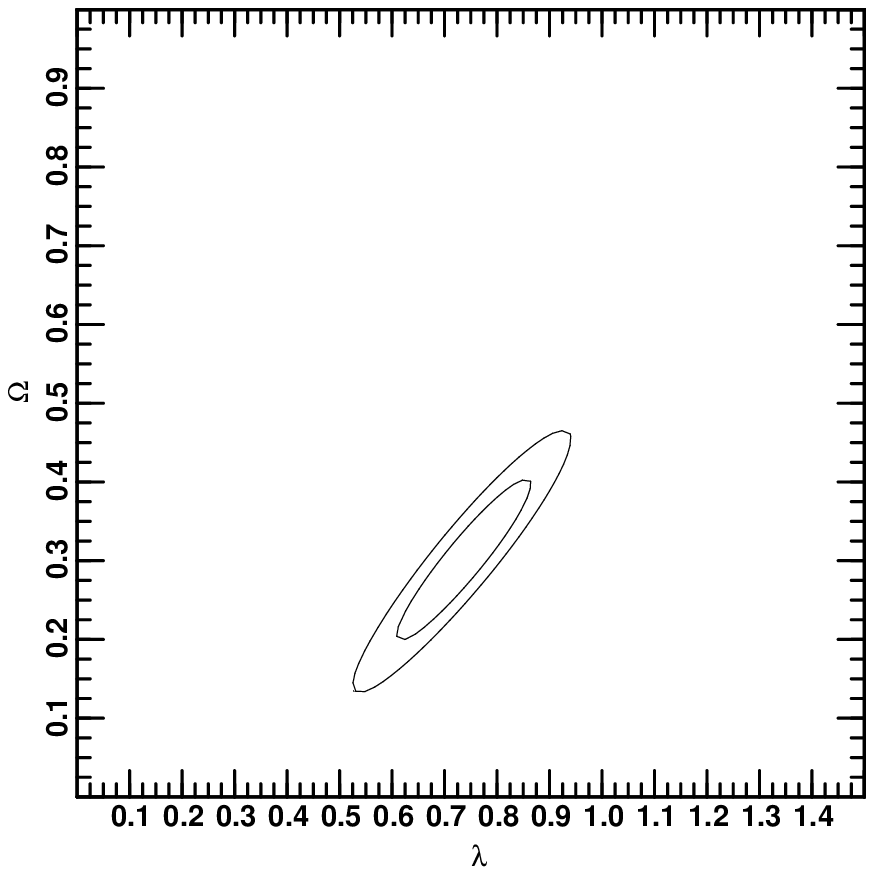,width=0.8\columnwidth}
\caption{Cut through the three-dimensional probability distribution
perpendicular to the $\eta$-axis for $\eta=0.955$.} 
\label{closeup-cut-3-lo}
\end{figure}
\ref{closeup-cut-3-lo} show cuts for $\eta = 0.005, 0.455, 0.955$.  The
contours become smaller and move to lower values of \lnull\ and \onull\
as $\eta$ becomes larger.  (Again, the contours at $0.954$ and $0.997$
cannot be distinguished in these plots.) \Fig~\ref{closeup-cut-3-lo} is
quite similar to standard presentations of the supernova constraints 
\citep[\eg][]{NSuzukietal2012a},
but keep in mind that these contours are a cut through the
three-dimensional contours for a fixed value of $\eta$, not
two-dimensional contours. If $\eta$ is substantially less than 1, then
not only is the allowed region much larger, but the `concordance model'
with $\lnull \approx 0.7$ and $\onull \approx 0.3$ is ruled out.
Qualitatively, this behaviour is easy to understand: there is some
degeneracy between $\eta$ and $\lnull + \onull$ since both increase the
amount of focusing in the beam, the former because there is more matter
in the beam and the latter because of the increase in the global
curvature, which is essentially $\lnull + \onull$.  When there is
essentially no matter in the beam, then the value of \onull\ is less
important and hence not as well constrained.  This means that $\lnull +
\onull$ can be realized \via\ a larger range of each parameter, making
the allowed region larger. The middle value of $\eta$ is that of the
global maximum probability.  (Since \lnull\ and \onull\ are better
constrained, the corresponding plots for fixed values of these
parameters, not shown here, are less interesting.) 

The `standard procedure' for reducing the number of parameters shown in
a plot is to \emph{marginalize} over the less interesting or `nuisance'
parameters.  This is shown in 
\begin{figure}
\epsfig{file=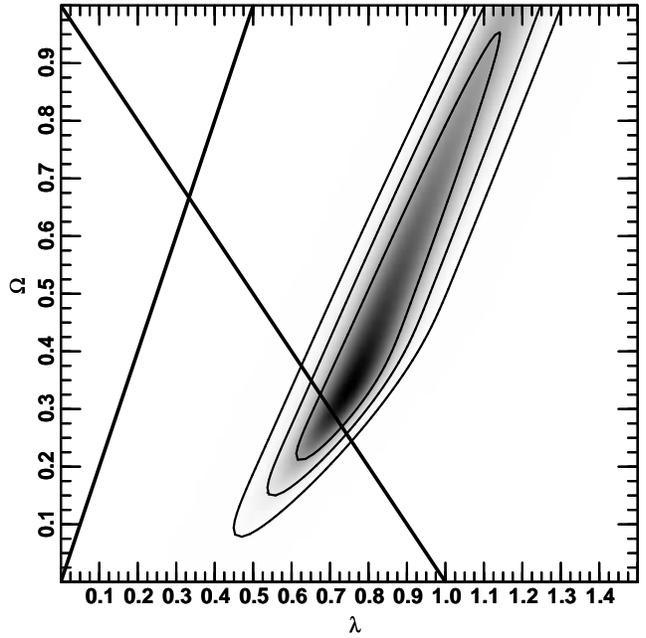,width=0.8\columnwidth}
\caption{Two-dimensional probability distribution obtained by
marginalizing over $\eta$.} 
\label{closeup-marginalize-lo}
\end{figure}
\Figs~\ref{closeup-marginalize-lo},
\begin{figure}
\epsfig{file=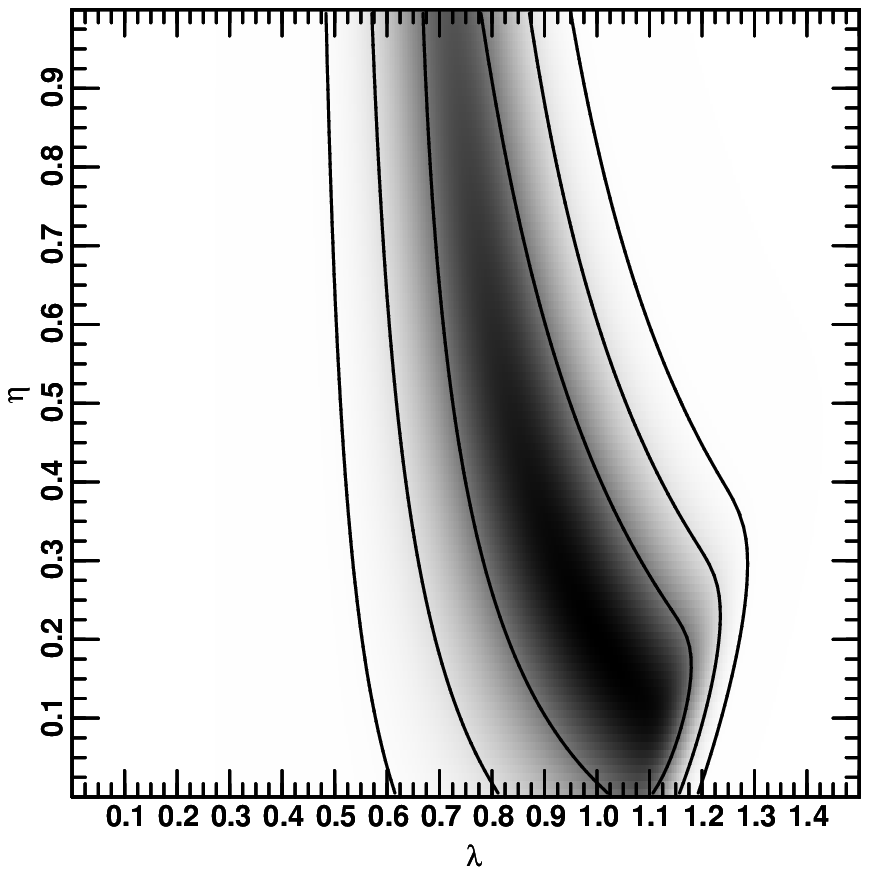,width=0.8\columnwidth}
\caption{Two-dimensional probability distribution obtained by
marginalizing over \onull.} 
\label{closeup-marginalize-le}
\end{figure}
\ref{closeup-marginalize-le}, and 
\begin{figure}
\epsfig{file=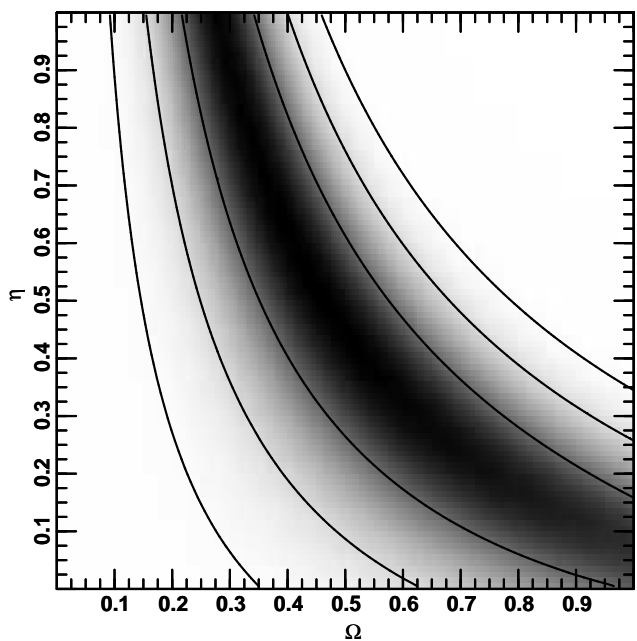,width=0.8\columnwidth}
\caption{Two-dimensional probability distribution obtained by
marginalizing over \lnull.} 
\label{closeup-marginalize-oe}
\end{figure}
\ref{closeup-marginalize-oe}.  Here, and in similar figures below, the
grey-scale corresponds to the probability.\footnote{It has become
fashionable to plot contours and have the regions between the contours
filled with a certain colour (or perhaps shade of grey).  This conveys
no information in addition to the contours themselves.  Of course, the
probability between two contours, or within the smallest contour, is not
everywhere the same, as is obvious from
\Fig~\ref{closeup-marginalize-lo}.  I have chosen to display this
potentially important information in addition to the contour curves.}
These are qualitatively similar to the projections.
\Fig~\ref{closeup-marginalize-lo} also contains two straight lines
corresponding to a flat universe with $\lnull + \onull = 1$ (negative
slope) and zero acceleration ($q_{0} = \frac{\onull}{2} - \lnull = 0$)
(positive slope).  Note that a flat universe is compatible with the data
but not required by them; in fact, the degeneracy in the constraints is
almost perpendicular to the flat-universe line.  In this particular
plot, the degeneracy corresponds roughly to $q_{0} \approx -0.6$; in
many of the other plots, the degeneracy in the \lnull--\onull\ plane is
closer to a constant value of $\onull - \lnull$ than to a constant value
of $q_{0} = \frac{\onull}{2} - \lnull$.  ($q_{0}$ was important
historically since the departure from the linearity of the $m$--$z$
relation at low redshift is proportional to $q_{0}$; nowadays quoting a
value for $q_{0}$ derived from the $m$--$z$ relation for higher-redshift
objects is neither necessary nor sufficient nor, in general,
meaningful.) 

Another approach is to \emph{maximize} the `nuisance' parameter, \ie\
for a given point in the plane of the plot, find the value of the third
parameter which maximizes the probability.  This is shown in 
\begin{figure}
\epsfig{file=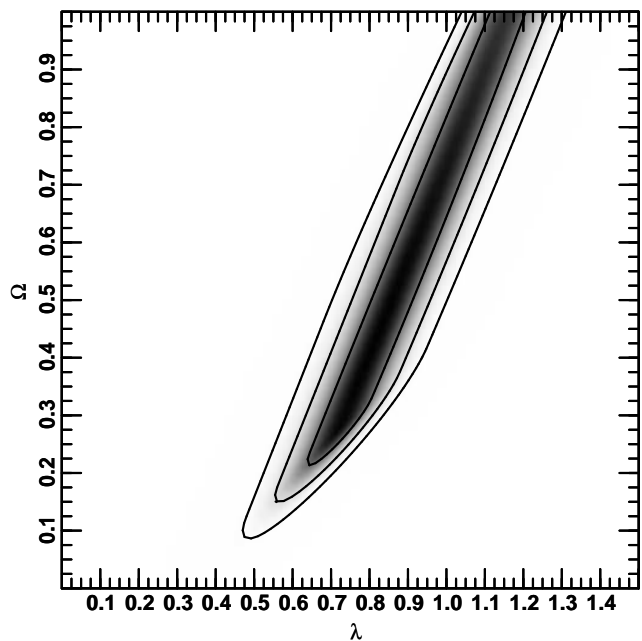,width=0.8\columnwidth}
\caption{Two-dimensional probability distribution obtained by maximizing
$\eta$.} 
\label{closeup-maximize-lo}
\end{figure}
\Fig~\ref{closeup-maximize-lo}.  (For these data, such plots are very
similar to those where the third parameter has been marginalized over,
so only this one example is shown.) 

Most discussion of the $m$--$z$ relation for Type~Ia supernovae has
concentrated not on contours of more than two dimensions, nor on some
reduction (projection, cut, marginalization, maximization) of these
higher-dimensional contours to two dimensions, but rather on
two-dimensional contours, \ie~with a $\delta$-function prior on the
nuisance parameters.  Almost always, of course, the (often implicitly
assumed) prior is $\eta=1$.  For comparison, in 
\begin{figure}
\epsfig{file=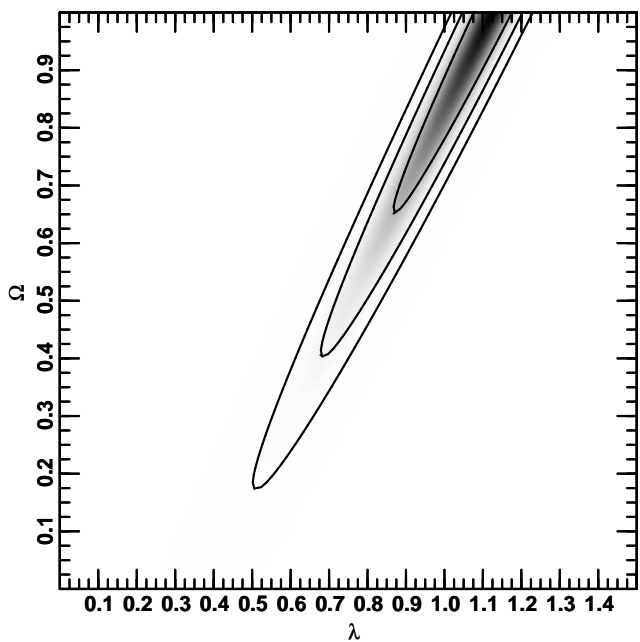,width=0.8\columnwidth}
\caption{Two-dimensional probability distribution for $\eta=0$.}
\label{closeup-cut-etazero-lo}
\end{figure}
\Figs~\ref{closeup-cut-etazero-lo}, 
\begin{figure}
\epsfig{file=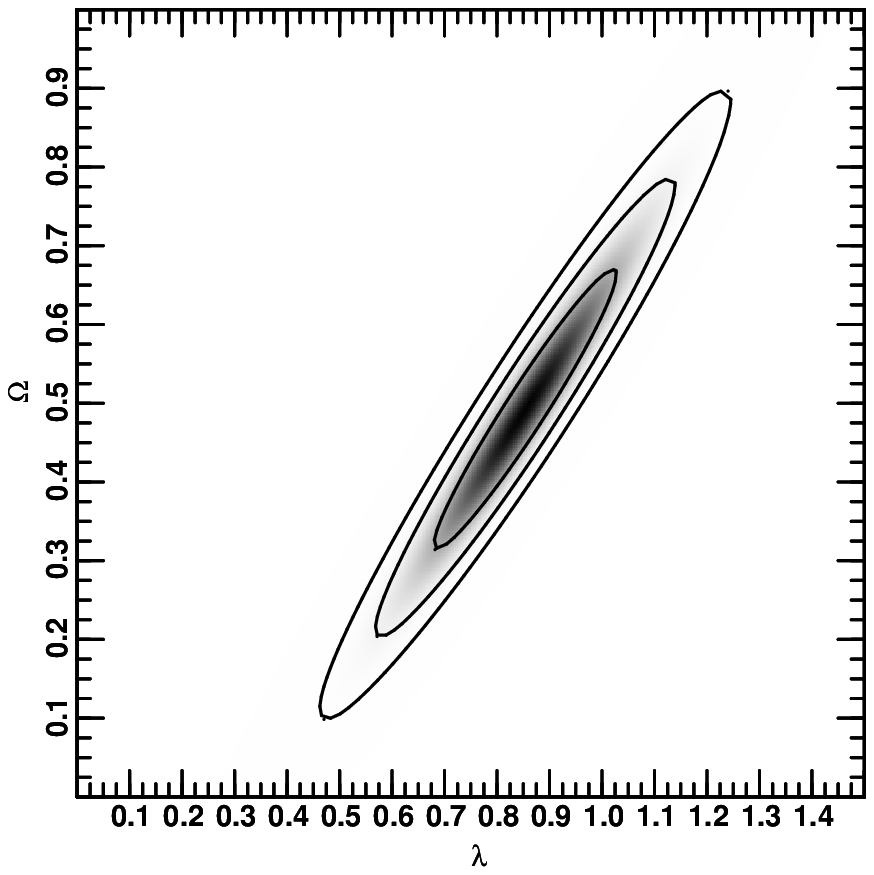,width=0.8\columnwidth}
\caption{Two-dimensional probability distribution for $\eta=0.455$.}
\label{closeup-cut-etabest-lo}
\end{figure}
\ref{closeup-cut-etabest-lo}, and
\begin{figure}
\epsfig{file=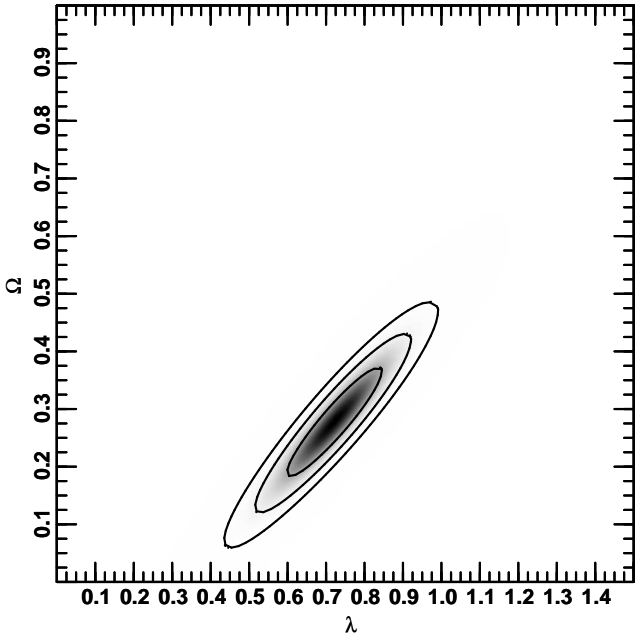,width=0.8\columnwidth}
\caption{Two-dimensional probability distribution for $\eta=1$.}
\label{closeup-cut-etaone-lo}
\end{figure}
\ref{closeup-cut-etaone-lo} I show constraints in the \lnull--\onull\
plane for fixed values of $\eta$, namely 0, 0.455 (the value at the
maximum of the three-dimensional probability distribution) and 1.  The
last should be compared with \eg~\fig~11 in 
\citet{MKowalski2008a}, 
but keep in mind that, as mentioned above, I have fixed \hnull\ and use
only the statistical uncertainties.  (See also \figs~1a and 5a in 
\citet{RAmanullaEMoertsellAGoobar2003a}.)  
Thus, \Fig~\ref{closeup-cut-etaone-lo} has slightly smaller contours
than similar plots elsewhere in the literature.  Again, this is
intentional so that any deviations from this fiducial plot (larger
and/or shifted contours) are due solely to the influence of $\eta$. 

Of course, little significance should be placed on variations in the
probability within the innermost contour, since the probability that the
point representing the true values of \lnull\ and \onull\ is only about
twice as likely to lie inside this contour than outside it.
Nevertheless, it is remarkable that the maximum of the probability in
\Fig~\ref{closeup-cut-etaone-lo} is at $\lnull =  0.7210938$ and $\onull
=  0.2773438$, \ie~at the values of the concordance model (within the
small uncertainties; these are much smaller than even the
$68.3$~\percent\ contour in
\Fig~\ref{closeup-cut-etaone-lo}).\footnote{For completeness, I quote
the exact position of the maximum as calculated on the grid; of course,
this does not imply that the maximum is known to greater precision than
the resolution of the grid.} Note that when fewer supernova data were
available, the \bestfit\ value was at much higher values of \lnull\ and
\onull; see \eg~\fig~1 in 
\citet{PHelbig99Ra}.  
(As mentioned above, the \bestfit\ value is often not visible in modern
versions of such plots, though of course it can be easily found in the
data used to make the plots.)  If the \bestfit\ value remains the same
when significantly more supernova data are available, then very probably
the true value will have been converged upon, even though the range of
values allowed, even at the $68.3$~\percent\ level, would include values
well outside what is acceptable when other cosmological constraints are
considered (\ie~joint constraints from several cosmological tests).
Normally, when more data are available one expects the new \bestfit\
value to be consistent with, but different from, the old \bestfit\
value, as has been the case with the supernova data up until now. 
However, looking towards the future, I don't expect the \bestfit\ values
for \lnull\ and \onull\ to change significantly, but do expect the
constraints from the supernova data to improve, which appears somewhat
puzzling.  A possible explanation for this is that the statistical
errors in the supernova data have been overestimated.  Note, however,
that the \bestfit\ values for the supernova data correspond to the
concordance model only if one assumes $\eta \approx 1$.  For $\eta =
0.455$, the concordance model lies very near the $95.4$~\percent\
contour, and for $\eta = 0 $ it is even outside the $99.7$~\percent\
contour.  The plots above illustrate that it is not possible to
appreciably constrain $\eta$ from the supernova data alone. However, the
fact that the supernova data suggest the concordance model only for high
values of $\eta$ could be seen as evidence that $\eta \approx 1$. 

A similar result is shown in 
\begin{figure}
\epsfig{file=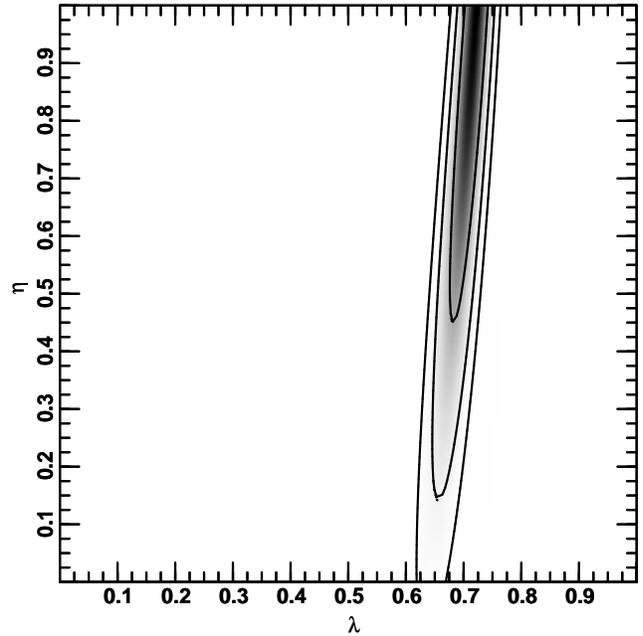,width=0.8\columnwidth}
\caption{Two-dimensional probability distribution for $k=0$.}
\label{closeup-cut-flat-le}
\end{figure}
\Fig~\ref{closeup-cut-flat-le}, where a flat universe ($\lnull + \onull
= 1$) has been assumed.  As in the other plots, \lnull\ is reasonably
well constrained, while $\eta$ is quite weakly constrained.  (In this
case, since $\onull = 1 - \lnull$, \onull\ is just as well constrained;
in general, \onull\ is less well constrained than \lnull.)  However,
note that the \bestfit\ value is for $\eta = 1$ and $\lnull \approx
0.72$; in other words, again the best fit is for the concordance model
with $\eta = 1$.  (This plot also shows the importance of plotting the
probability and not just a few contours.) 

To illustrate the change in the effect of $\eta$ now that more supernova
data are available, 
\begin{figure}
\epsfig{file=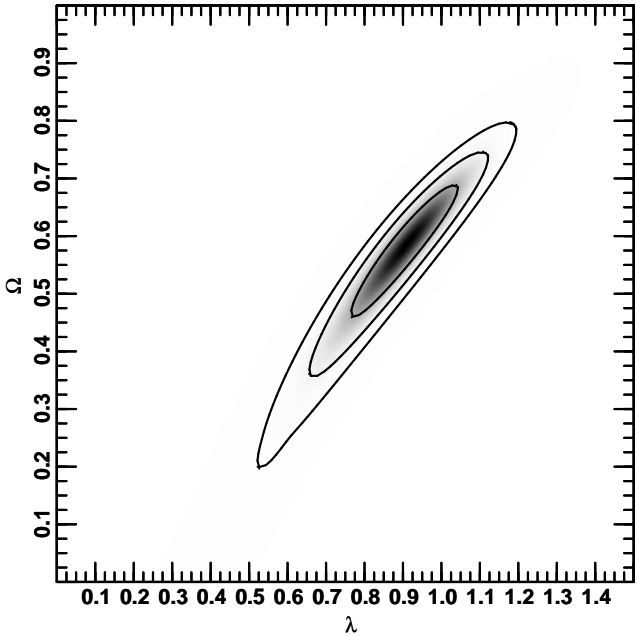,width=0.8\columnwidth}
\caption{Two-dimensional probability distribution with $\eta=f(\onull)$.}
\label{closeup-cut-vareta-lo}
\end{figure}
\Fig~\ref{closeup-cut-vareta-lo} shows the constraints where $\eta$ is a
function of \onull, namely $\eta = 0$ for $\onull \le 0.25$ and $1 -
0.25$ for $\onull \ge 0.25$.  This should be compared with \fig~8 in 
\citet{SPerlmutteretal99a}.  
In that figure, the red contours were calculated in the same way as
those in \Fig~\ref{closeup-cut-vareta-lo}.  In the same figure, the
green contours were calculated in the same way as in
\Fig~\ref{closeup-cut-etazero-lo}. The comparison illustrates vividly
the fact that the effect of $\eta$ can no longer be neglected.  While 
\citet{SPerlmutteretal99a} 
concluded that, at least in the interesting part of parameter space, the
constraints on \lnull\ and \onull\ from the supernova data did not
depend heavily on the assumed value of $\eta$, this is definitely no
longer the case. 

While the supernova data cannot usefully constrain $\eta$, as has been
shown above, the fact that they result in the concordance model if one
assumes $\eta \approx 1$ suggests that $\eta \approx 1$.  Since there
are many cosmological tests completely independent of the supernova
data, and also independent of the value of $\eta$, which suggest the
concordance model (this is of course why it is called the concordance
model), one can assume the concordance values for \lnull\ and \onull\
and calculate the probability of $\eta$ from the supernova data with
these additional constraints; this is shown in 
\begin{figure}
\epsfig{file=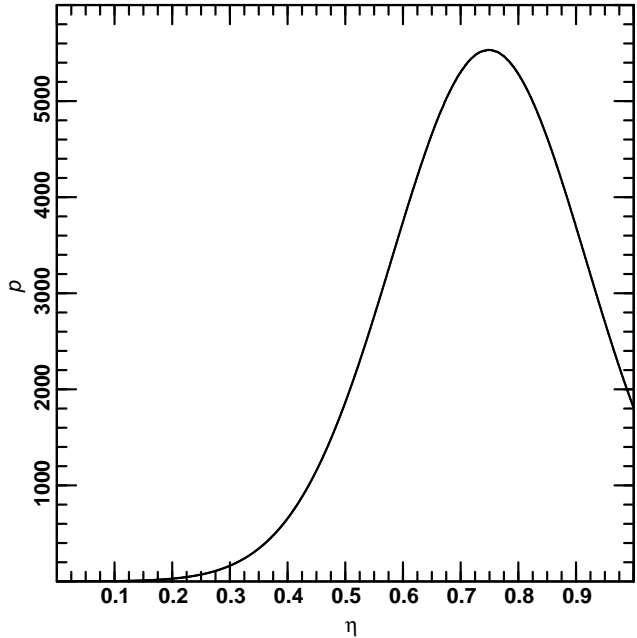,width=0.8\columnwidth}
\caption{One-dimensional probability distribution for the concordance
model.} 
\label{closeup-cut-concordance-e}
\end{figure}
\Fig~\ref{closeup-cut-concordance-e}.  The \bestfit\ value is $\eta =
0.7485$ while the formal statistical limits are: 
\begin{eqnarray*} 
0.60 < \eta < 0.90\ (68.3\%) \\ 
0.46 < \eta < 1.00\ (95.4\%) \\ 
0.28 < \eta < 1.00\ (99.7\%) 
\end{eqnarray*} \quad .
%
\begin{figure}
\epsfig{file=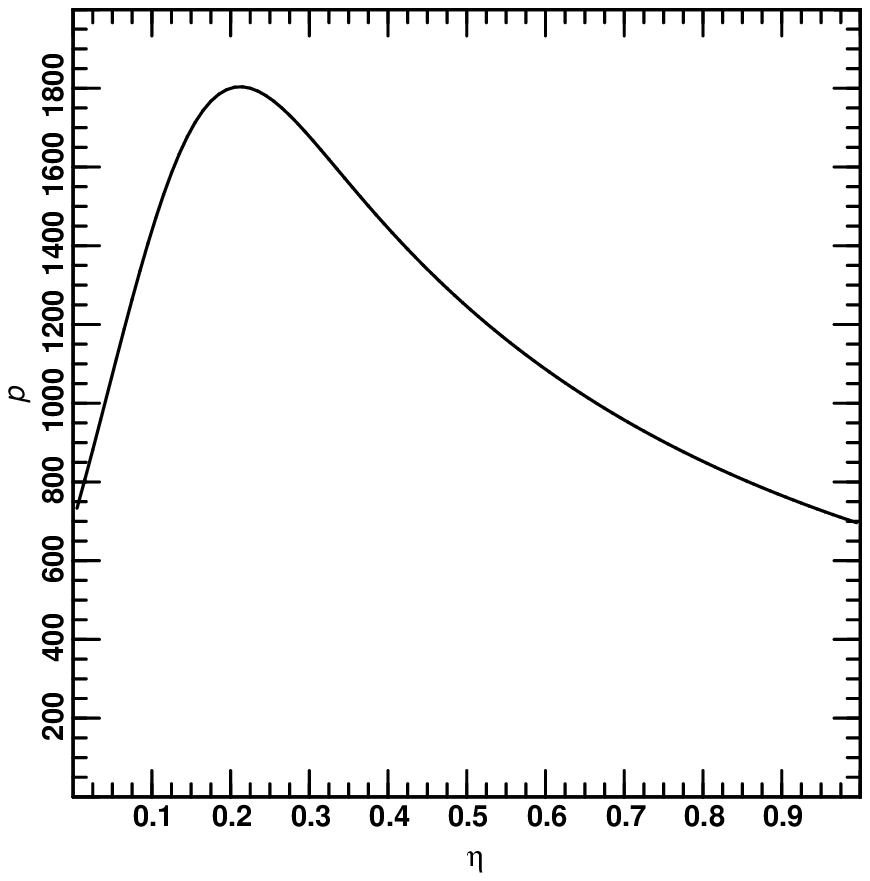,width=0.8\columnwidth}
\caption{One-dimensional probability distribution after
mar\-gi\-na\-li\-zing over \lnull\ and \onull.} 
\label{closeup-marginalize-e}
\end{figure}
(This can be contrasted with \Fig~\ref{closeup-marginalize-e} which
shows the value of $\eta$ preferred by the supernova data alone; \lnull\
and \onull\ have been marginalized over.)  While $\eta = 1$ is not ruled
out at high confidence, lower values of $\eta$ are ruled out at a high
level of statistical significance.\footnote{This can be contrasted with
the location of the maximum of the three-dimensional probability
distribution, where the \bestfit\ values are $\lnull = 0.8609375$,
$\onull = 0.5015625$, and $\eta = 0.455$.  While this lies outside the
allowed region of parameter space as determined from cosmological tests
other than the $m$--$z$ relation for Type~Ia supernovae, the allowed
region is quite large and the concordance model with $\eta = 1$ is
within the $68.3$~\percent\ contour.  Even though the constraints on
\lnull\ and \onull\ are of course weaker if $\eta$ is allowed to vary, a
significant portion of the three-dimensional parameter space can be
ruled out, and portions of the \lnull--\onull\ plane are also ruled out,
though no additional region is ruled out which is not already ruled out
by other cosmological tests.} This suggests that $\eta$ is relatively
large, even though the beams of supernovae at cosmological distances are
extremely thin and this cosmological test should suggest a value of
$\eta$ lower than that of any other cosmological test known today.  One
would not expect to obtain $\eta=1$ since some matter is associated with
galaxies which are outside the beam; such mass contributes about $0.1$
to \onull. \Fig~\ref{closeup-cut-concordance-e} thus suggests that dark
matter is distributed much more smoothly than galaxies.  While the beam
of a supernova at cosmological distance is almost a fair sample of the
Universe, it is an even fairer sample of dark matter.  Dark matter is
thus not significantly clumped at the scale of a supernova beam.

\section{Summary, conclusions and outlook}
\label{final}

The following conclusions were more or less expected.
\begin{itemize}
\item[(i)] Constraints on \lnull\ and \onull\ are weaker if $\eta$ is not
      constrained. 
\item[(ii)] The concordance model is reasonably probable.
\item[(iii)] There is a degeneracy between $\eta$ and the amount of spatial
      curvature ($\lnull + \onull$). 
\item[(iv)] \lnull\ is constrained best, then \onull, then $\eta$.
\end{itemize}
The following conclusion was neither expected nor surprising.
\begin{itemize}
\item[(i)] Even when $\eta$ is allowed to be a free parameter, the $m$--$z$
      relation for Type~Ia supernovae is not compatible with
      $q_{0} = \frac{\onull}{2} - \lnull \geq 0$, and thus implies that 
      the Universe is currently accelerating.\footnote{%
\citet{EMoertsellClarkson2009a} 
      have shown that this conclusion also holds for a much wider class 
      of models than the Friedmann\ndash Lema\^{\i}tre models considered 
      here.}  (Even though the $m$--$z$ relation for Type~Ia supernovae 
      is one of the key pieces of evidence supporting the cosmological 
      `concordance model' with $\lnull \approx 0.7$ and $\onull \approx 
      0.3$, it is not an essential piece in the sense that combinations 
      of other tests still result in the same concordance model.  
      Nevertheless, it is still an important piece of evidence in favour 
      of the concordance model since it is the only single test which, 
      without additional assumptions, implies $q_{0} < 0$, \ie\ a 
      Universe which is currently accelerating.)
\end{itemize}
The following conclusions are somewhat surprising.
\begin{itemize}
\item[(i)] The overall (in the three-dimensional parameter space) \bestfit\ 
      values for \lnull\ and \onull\ are ruled out by other cosmological
      tests.  Probably, this \bestfit\ point is the result of
      overfitting: its probability is not significantly higher than
      elsewhere and the allowed region is quite large. 
\item[(ii)] If one \emph{assumes} $k=0$ then the best fit is very close to the
      concordance model \textit{and} has $\eta=1$. 
\item[(iii)] If one \emph{assumes} $\eta=1$, then the best fit is very close to
      the concordance model. 
\item[(iv)] If one \emph{assumes} the concordance model, then one can probably 
      rule out low values of $\eta$, even though the relevant scale is 
      extremely small, which implies that dark matter is much less 
      clustered than galaxies are. 
\item[(v)] We cannot rule out $\eta=1$, and there is some tentative evidence
      for it. 
\end{itemize}

To summarize, allowing $\eta$, which is otherwise only weakly
constrained, as a free parameter significantly alters both the best fit
in the \lnull--\onull\ plane and the allowed region of this plane.  The
concordance model is, however, still allowed.  There are hints that
$\eta \approx 1$, though these are not statistically significant when
examined in the three- or two-dimensional parameter space.  On the other
hand, if one assumes the concordance values for \lnull\ and \onull, low
values of $\eta$ can probably be ruled out, which is not obvious
considering the very small scales involved; this implies that dark
matter is very homogeneously distributed. 

One might have thought that the increase in the number of data points
since 
\citet{SPerlmutteretal99a} 
would allow some sort of useful constraint to be placed on $\eta$ from
the supernova data without further assumptions.  This is not the case.
Even worse, if $\eta$ is allowed to vary, then the conclusions about the
cosmological model derived from the $m$--$z$ relation for Type~Ia
supernovae are not as robust.  However, as discussed in
\sect~\ref{intro}, current constraints from combinations of cosmological
tests without using the supernova data determine the `concordance model'
with $\lnull \approx 0.7$ and $\onull \approx 0.3$ to rather high
precision.  It is thus perhaps more interesting to assume the
concordance model and use the supernova data to constrain $\eta$,
especially since $\eta$ is otherwise difficult to measure.  Indeed, as
shown in \Fig~\ref{closeup-cut-concordance-e}, current data already
provide interesting constraints.  It is also extremely interesting that
the supernova data have the \bestfit\ values for \lnull\ and \onull\
corresponding to those of the concordance model if and only if $\eta
\approx 1$ is assumed.  (Note that while the \bestfit\ value of $\eta$
assuming the concordance model is $\approx 0.75$, the \bestfit\ values
of \lnull\ und \onull\ assuming $\eta \approx 0.75$ are different from
those of the concordance model.) If this is not a statistical fluke, it
could indicate that $\eta \approx 1$, which is somewhat surprising since
the value of $\eta$ as `felt' by the supernova might be expected to be
somewhat less, because the corresponding beams are extremely thin.  The
fact that even the supernova data `want' $\eta \approx 1$ could indicate
that dark matter is distributed extremely homogeneously.  See 
\citet{DHolz98a} 
for a different expression of the same idea.  Alternatively, this could
be evidence that the `Safety in Numbers' scenario mentioned in
\sect~\ref{theory} is in fact a valid approximation. 

In contrast to the first useful determinations of \lnull\ and \onull\
from the $m$--$z$ relation for Type~Ia supernovae 
\citealp[\eg][]{PGarnavichetal1998a,ARiessetal98a,SPerlmutteretal99a}, 
where the effect of $\eta \neq 1$ had a negligible effect on the
constraints derived, at least in the `interesting' region of the
\lnull--\onull\ parameter space, with the larger number of supernovae now
available, this is no longer the case. At the same time, current
supernova data alone cannot usefully constrain $\eta$ (though this might
be possible if other cosmological data are taken into consideration, as
discussed in the previous paragraph).  This should be taken into account
in attempts to determine further parameters, such as $w$, the
equation-of-state parameter for dark energy.  When more supernova data
become available, especially at higher redshift, it might be possible to
usefully constrain $\eta$ and/or discriminate between the effect of
$\eta$ and other parameters such as $w$.  (The difference in apparent
magnitude for different values of $\eta$ increases with increasing
redshift, while the difference due to different values of \lnull\ and
\onull\ is stronger (than that due to variation in $\eta$) at lower
redshift and, for some sets of models, decreases at higher redshift.) 
While allowing $\eta$ to be a free parameter, but constant as a function
of redshift and for different lines of sight, is certainly not the last
word with respect to the influence of locally inhomogeneous cosmological
models on the $m$--$z$ relation for Type~Ia supernovae, it does
demonstrate that care is needed when interpreting conclusions derived
from assuming $\eta = 1$.  At least, the uncertainty in \lnull\ and
\onull\ must be correspondingly increased.  While it might be possible
to decrease this with a more realistic model, it is no longer possible
to assume $\eta = 1$ and have confidence in the parameters and their
uncertainties resulting from an analysis of the $m$--$z$ relation for
Type~Ia supernovae.

\section*{\ack}

I thank Nils Bergvall, Phil Bull, Bruno Leibundgut, and an anonymous
referee for helpful comments.  Figures were produced with the
\software{gral} software package written by Rainer Kayser.

\input{etasnia.bbl}

\bsp
\label{lastpage}
\end{document}